\documentstyle[epsfig,aps,prl,twocolumn]{revtex}

\begin{document}

\tightenlines
\wideabs{

\title{Nuclear Inelastic X-Ray Scattering of FeO to 48 GPa}

\author{Viktor V. Struzhkin, Ho-kwang Mao, Jingzhu Hu, 
 Markus Schwoerer-B\"ohning, Jinfu Shu, and Russell J. Hemley}
\address{Geophysical Laboratory and Center for High Pressure Research,
Carnegie Institution of Washington,\\
5251 Broad Branch Road NW, Washington, DC 20015, USA, and\\
HPCAT, Advanced Photon Source, Argonne National Laboratory,
Argonne, IL 60439, USA}

\author{Wolfgang Sturhahn, Michael Y. Hu, and Ercan E. Alp}
\address{Advanced Photon Source, Argonne National Laboratory,
Argonne, IL 60439, USA}

\author{Peter Eng and Guoyin Shen}
\address{University of Chicago, 5640 South Ellis,
Chicago, IL 60637, USA, and 
GSECARS, Advanced Photon Source, Argonne National Laboratory,
Argonne, IL 60439, USA }

\date{\today}
\maketitle

\begin{abstract}
The  partial density of vibrational states has been measured for Fe 
in compressed FeO (w\"ustite) using nuclear resonant inelastic
 x-ray scattering. Substantial changes have been observed in  the overall shape 
of the density of states close to the magnetic transiton around 20 GPa from 
the paramagnetic (low pressure) to the antiferromagnetic (high pressure) state. 
Our data indicate a substantial softening of the aggregate sound velocities far below the 
transition, starting between 5 and 10 GPa. This is consistent with recent radial x-ray 
diffraction measurements of the elastic constants in FeO. 
The results indicate that  strong magnetoelastic coupling in FeO  is 
the driving force behind the changes in the phonon spectrum of FeO. 
 
\end{abstract}

}

\narrowtext

The  study of the  electronic  and  magnetic  properties of   simple
transition-metal compounds is an important topic in  diverse
fields  ranging from solid-state physics to Earth  sciences. Iron
oxide FeO (w\"ustite)  belongs to the group  of highly correlated
transition  metal oxides,  being an archetypal insulating
antiferromagnetic material at zero temperature. NiO, CoO, MnO fall in the same 
group of materials, which are still not  well understood by theory. 
They also present a challenge to experimentalist because the predicted 
high pressure transformations in these materials  (metallization, high-spin to  
low-spin  transitions) may occur at extreme pressures ( e. g. $>$ 100 GPa).  
FeO stands out from this group 
of materials in the sense that it is a possible major
constituent of the Earth's lower mantle and core, thus its pressure
and temperature  dependent properties are very important  for the 
understanding of the Earth's interior. Here we present a study
of  the vibrational density of states of  isotope-enriched
Fe$_{0.947}$O using high-resolution  nuclear resonant  inelastic x-ray
scattering. We observe changes in the density of states  that
are consistent with the softening of the aggregate shear sound wave velocity
of  w\"ustite under pressure \cite{Mao96}, which we associate with the phase
transition from the cubic  paramagnetic phase to the rhombohedrally distorted
antiferromagnetic phase stable at higher pressures. We show that the phonon density 
of states is affected by magnetoelastic coupling between phonon and spin subsystems 
in FeO.

Experiments were performed at the synchrotron beamline SRI-CAT 3ID of the 
Advanced Photon Source (APS), Argonne National Laboratory.
The details of the diamond cell and the experimental setup are reported 
elsewhere \cite{Mao00}.  The sample was $\sim$ 25 $\mu$m in diameter 
(loaded without pressure medium), 
and diamonds with flat 400 $\mu$m culets were used. The maximum 
counting rate in the phonon wing ranged from 10 cps at 0.9 GPa 
to 2 cps at 48 GPa. The measured spectra, i. e., count rate as a function 
of the energy difference between the incident photon energy and the nuclear 
transition energy, $I(\Delta E)$, were converted to phonon DOS profile according to 
the data analysis procedure described by Hu et al. \cite{Hu99}. 
Typically, 10-20  DOS spectra (one hour/spectrum) at the same pressure 
were summed together. 
The pressure was measured using the ruby fluorescence line and 
a nonhydrostatic pressure scale \cite{Maop}.

\begin{figure}
\centerline{\epsfig{file=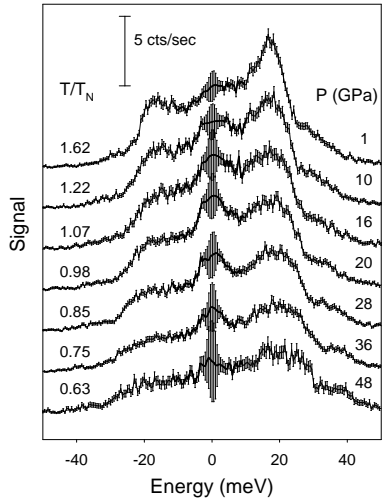,width=6cm}}
\caption{Inelastic part of the signal in FeO as a function of pressure  
(elastic peak is subtracted). The region close to zero energy transfer is shown 
using bold lines. Note enhanced density of states close to T/T$_N$ $\sim$ 1. 
Ratio  T/T$_N$ was estimated using T$_N$=198 K at P=0.1 MPa, 
and data from Ref.[7].} 
\label{fig1}
\end{figure}

\begin{figure}
\centerline{\epsfig{file=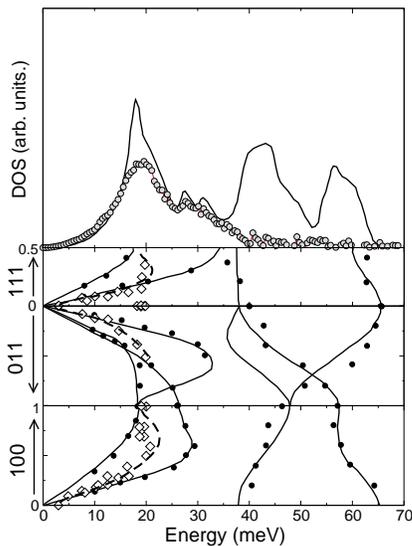,width=6cm}}
\caption{Dispersion of phonons and magnons in FeO from neutron inelastic
scattering experiments. Upper part shows comparison of phonon DOS 
calculated from neutron measurements (solid line) with our partial 
DOS for iron at 0.9 GPa (grey circles), 
300 K. Also shown are measured and calculated phonon (circles and solid lines) and magnon 
(diamonds and dashed lines) branches from Ref.[5]. }
\label{fig2}
\end{figure}

Experimental data are presented in Fig.\ref{fig1}. In Fig.\ref{fig2} we  compare the phonon 
DOS of FeO as derived from neutron scattering \cite{Kugel77} and the partial phonon 
DOS of Fe at 0.9 GPa from our measurements. High frequency oxygen 
modes are absent from our data, 
indicating that Fe vibrational amplitude is negligible at oxygen vibrational frequencies. 
One apparent feature of the experimental 
data is an inelastic peak at small energy transfers which develops at 10 GPa, and persists 
through almost the entire pressure range of the present experiment to 
48 GPa. It is most pronounced
near  16 and 20 GPa,  which is exactly the pressure range of the reported transition
from the paramagnetic rock-salt-type structure to the antiferromagnetic rhombohedral structure 
\cite{Yagi85,Fei94}.  The transition is very sensitive to (nonhydrostatic) stress conditions
\cite{Yagi85}; for example, the splitting of the (111) diffraction line being smeared 
between 10 to 18 GPa 
if no pressure medium was used, and occuring around 16 GPa \cite{Yagi85} 
to 17 GPa \cite{Mao96} under quasihydrostatic conditions. 

\begin{figure}
\centerline{\epsfig{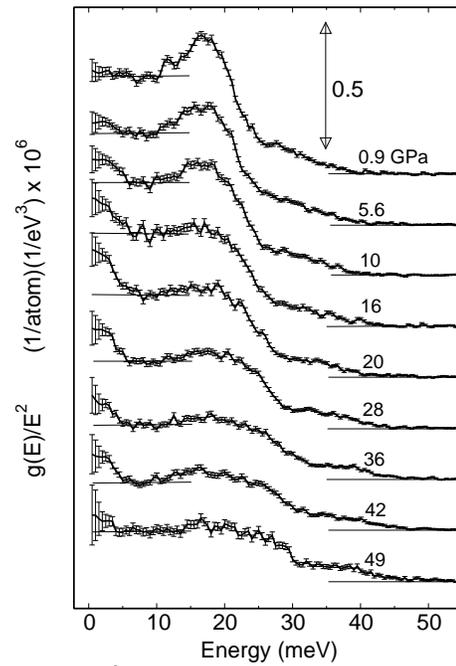}}
\caption{g(E)/E$^2$ derived from partial DOS  in FeO  
as a function of pressure. For the Debye model, the lower energy part should be 
a horizontal straight line.}
\label{fig3}
\end{figure}

The calculated partial density of 
states weighted by the square of the energy  E$^2$ is plotted versus energy in  Fig.\ref{fig3}. 
In Fig.\ref{fig3}, the density of states in the Debye approximation 
would be a straight horizontal line, related to the aggregate 
sound velocity of the material \cite{Mao00,Lubers}. The apparent feature of our data 
is the pronounced softening of the low energy vibrational spectrum. The resolution of the present 
measurements (2.4 meV) 
does not allow us to follow the softening to the low frequency region, where most static 
and ultrasonic measurements are performed. However, this behavior suggests   large effects 
on the static elastic constants at the N\`eel transition. 

The temperature dependence of the elastic constants of Fe$_{0.92}$O at ambient 
pressure was investigated in detail  by Sumino et al. \cite{Sumino80}. 
Substantial softening of the shear 
constant C$_{44}$ was found close to the N\`eel transition. Similar anomalies have also been  
observed  in MnO (softening of C$_{44}$) \cite{overview1}. 
The magnetic structure studies \cite{Shull,Li55,Roth58} by neutron scattering
have shown that the Mn spins align ferromagnetically within a given (111) plane, 
and these planes are stacked antiferromagnetically in the $<$111$>$  direction. 
Similarly, FeO has the same magnetic structure, with magnetic 
moments pointing in the $<$111$>$ direction. 
At temperatures below the N\`eel temperature MnO undergoes a lattice contraction along the 
antiferromagnetic  stacking direction $<$111$>$ \cite{Morosin70}. The distortion in FeO has 
different sign, the elongation along the body diagonal. The magnetic system 
breaks into multiple domains (called T domains)
corresponding to the antiferromagnetic stacking along the four possible $<$111$>$ 
directions within the crystal. Experimental \cite{Morosin70,Bloch73,Pepy74} and theoretical  
\cite{Lines65,Bartel70,Seino82} studies in MnO invoke exchange-striction effects to 
explain the character of the transition. 
The results of these studies can be summarized 
as follows. The energy balance is determined by minimizing the sum of magnetic 
exchange energy and elastic energy, resulting in a rhombohedral distortion

\begin{equation}
\Delta=z_1NJ_1\epsilon_1 S^2 /24C_{44} 
\label{eq:rhdist}  
\end{equation}

\noindent Here $\Delta$ denotes the distortion of  trigonaly deformed cube corner 
angles ${1\over 2}\pi \pm \Delta$, $C_{44}$ is the shear constant,
$J_1$ is the exchange integral for the nearest neighbors, $z_1$ is the number 
of the nearest neighbors, $\epsilon_1=-r\delta lnJ_1/\delta r$, and 
$S^2=<S_i \cdot S_j>_{nn}^p-<S_i \cdot S_j >_{nn}^a$ is a correlation function 
for nearest neighbors with 
parallel spins (residing within the same $<$111$>$ plane)  and antiparallel  spins in neighboring  
$<$111$>$ planes.  The numerical value of $\Delta$ is about 0.5 degree at 4 K in MnO. 
Actually, according to Bartel, \cite{Bartel70} $\Delta$ is proportional to the sublattice 
magnetization and is a sensitive measure of the order parameter of the magnetic phase.  

However, FeO and CoO have much larger volume anomalies below T$_N$ and their 
magnetic properties according to Kanamori \cite{Kanamori1,Kanamori2} 
cannot be described by the theory developed by Lines and Bartel \cite{Bartel70}. 
In a crystal field of cubic symmetry, the 
orbital degeneracies of Fe$^{2+}$ and Co$^{2+}$ are not completely removed, and the residual 
orbital angular momenta contribute to the energy balance through spin-orbit coupling and 
the direct effect of orbital magnetic moments, resulting in observable 
magnetostriction effects. Substantial contribution  to 
the magnetostriction effects in FeO is due to the magnetic anisotropy energy \cite{Kanamori2}. 
Kanamori \cite{Kanamori2} calculated the equlibrium strain components for FeO resulting 
from magnetostriction.  He used the general formulation derived by Kittel 
for cubic systems \cite{Kittel} (all $C_{ij}$'s  are cubic elastic constants, $B_1$ and $B_2$ 
are cubic magnetoelastic constatnts):

\begin{eqnarray}
\nonumber e_{ii}=B_1 \left( (1/3)-\alpha_i^2 \right) /(C_{11}-C_{12}), \\ 
e_{ij}=-B_2 \  \alpha_i \alpha_j/C_{44} 
\label{eq:strain} 
\end{eqnarray}

\noindent where $\alpha_i$ are direction cosines of the magnetization. 
Kanamori's treatment leads to an elongation along [111] diagonal in FeO, which agrees 
with  experimental results \cite{Roth58}. From the experiments on FeO at 95 K \cite{Rooksby51} 
one derives  a rhombohedral angle of 59$^\circ$ 32', or e$_{xy}$=0.705 \%. To estimate  
the rhombohedral distortion resulting from the exchange interaction (Eq.\ref{eq:rhdist}) in FeO, 
we need reliable information about the exchange integrals and their dependence on the lattice parameter,
which to our knowledge is not available at the moment.
The experimental values of distortion \cite{Mao96,Yagi85} seem to be quite close to the Kanamori's 
calculation. Moreover, rhombohedral distortion is enhanced at high pressures 
\cite{Yagi85}, indicating increasing  B$_2$ in FeO under pressure.

The  cubic constant $B_2$ is responsible for the magnetoelastic 
coupling between the phonon and the magnon branches, and our results on the enhanced phonon 
density of states are directly related to the magnetoelastic coupling in FeO. 
The expression for phonon dispersion including the magnetoelastic 
coupling follows from the equations of motion for the magnetic 
moment \cite{spinwave,expand}: 

\begin{equation}
(\omega^2-\omega_m^2)(\omega^2-\omega^2_s)-{{g k^2 B_2^2 \Omega}\over{2 \rho M_0 }}=0
\label{eq:couplmgph}
\end{equation} 
  
\noindent Here $\omega_s$ is the frequency  of sound wave, and 
$\omega_m^2=\Omega g M_0(\beta +\alpha k^2)$ is the spin wave  frequency 
in the absence of magnetoelastic coupling; $\rho$ is the density  of the material, 
$M_0$ is the  magnitude of  the magnetic moment of one sublattice, 
$\Omega = gM_0(\beta+2\gamma)$.   
The parameters $\alpha$, $\gamma$, and $\beta$ are related to the exchange interactions  
and magnetic anisotropy in the material \cite{spinwave},  $g$ is the gyromagnetic ratio, 
and $k$ is a wavevector. 

The calculated dispersion relations for FeO at 28 GPa with magnetoelastic coupling 
included \cite{param} are shown in  Fig.4a (estimated at  T=295 K), 
corresponding sound velocities are shown in Fig.4b.  
Dispersion relations  (Fig.4a) change by less than 1 meV, well within the resolution of 
neutron inelastic scattering experiment. However, as is evident from Fig.4b, the effect on 
the sound velocities 
is more pronounced, being almost 20-30 \% within the energy transfer range up to 5 meV. 
This agrees reasonably well with the enhancement of the 
density of states which we observe in our experiment (Fig.3). Quantitave agreement may be sought  
along the lines of more elaborated theoretical models, similar to Ref.\cite{Gann}. 

At present we do not have enough experimental information and 
theoretical understanding to better constrain  
the magnetoelastic coupling  at T/T$_N$ $\sim$ 1. However, as follows from our observations,
the effect is most pronounced close to the N\'eel transition. Our observations support the 
M\"ossbauer measurements, where the magnetic transition was observed starting from 8 GPa 
under nonhydroctatic conditions. The reason is that uniaxial strain along the body diagonal 
may induce magnetic moments well below the transition point determined under  hydrostatic 
conditions (17 GPa). This is consistent with the notion that 
this second order phase transition is smeared by the external field (uniaxial strain), which 
is proportional to the order parameter of the broken symmetry phase \cite{Landau}.

In summary, we have observed substantial softening of the density of states at energy 
transfers below 10 meV  in FeO at pressures close to 15-20 GPa, which persists in its 
antiferromagnetic phase up to 40-48 GPa. We relate the observed softening to the effect 
of the magnetoelastic coupling in this material. 
The theoretical estimates \cite{Kanamori2,Bartel70} show that both rhombohedral distortion 
and magnetoelastic coupling include substantial contributions from the magnetic anisotropy 
energy in FeO, which arises from the spin-orbit coupling in the orbitally degenerate ground state 
\cite{Kanamori1,Kanamori2,Solov98}. Further experimental work is required to clarify the 
effect of exchange-driven magnetoelastic contribution.

Portions of  this  work was  performed  at GSECARS, which  is
supported by the NSF, W.M. Keck  Foundation and the  USDA.  The APS is
supported by the DOE under Contract No.  W-31-109-Eng-38.
V. V. S. acknowledges Ron Cohen for bringing to his attention Ref.\cite{Solov98}.

\begin{figure}
\centerline{\epsfig{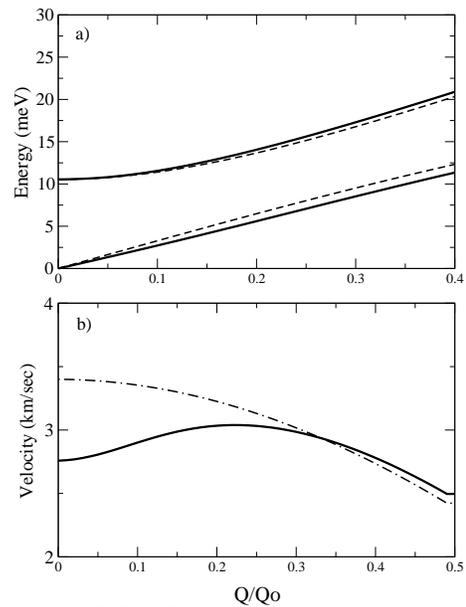}}
\caption{Model for the magnetoelastic coupling in FeO. a) The 
interacting transverse phonon and magnon branches. Noninteracting 
bare frequencies are shown with dashed lines, the dispersion branch 
for magnons is calculated according to [31], the phonon 
dispersion was approximated by $E(Q)=2Q_0 v_s/\pi sin((\pi/2)(Q/Q_0))$ 
using the sound velocity 3.4 km/sec at Q=0. 
b) Calculated sound velocity including magnetoelastic coupling (solid line) 
and without magnetoelastic coupling (dash-dotted line).} 
\label{fig4b}
\end{figure}


\begin{references}
\bibitem{expand} We have used Eq.12 from ref.\cite{spinwave} for zero external magnetic field, 
we neglect damping and eddy current contributions, and we use the relation 
$\delta_1-\delta_2=B_2/2M_0^2$ for magnetoelastic constants 
(see Ref.\cite{spinwave} for $\delta_1$ and $\delta_2$ notation).  
Magnetoelastic waves are assumed to travel along the body diagonal of the cubic structure of FeO.
\bibitem{param} We estimated $B_2$=-2.5 10$^{10}$ erg/cm$^{3}$ from 
the rhombohedral distortion using Eq.2; 
$\rho$=6.2 g/cm$^3$, we assumed $\gamma \ll \beta$, and  estimated $\alpha$=10$^{-6}$ erg$\cdot$cm 
from P=0.1 MPa data \cite{Kugel77}.
M$_0$=1400 Gauss (mean field estimate for $T/T_N$=0.85, 
$\mu_{eff}$=5.1$\mu_B$ \cite{landolt}); 
$\Omega$=$gM_0\beta$=2.6 THz (11 meV) was estimated from Raman data \cite{raman}.
\end{references}
\end{document}